\documentclass[preprint,showpacs,preprintnumbers,amsmath,amssymb]{revtex4}

\usepackage{graphicx}
\usepackage{dcolumn}
\usepackage{bm}

\begin{document}

\preprint{APS/123-QED}

\title{Nernst Effect as a Probe of Local Kondo Scattering in Heavy Fermions}

\author{$^{1}$Peijie Sun}
\author{$^2$Frank Steglich}
\affiliation{%
$^1$Beijing National Laboratory for Condensed Matter Physics, Institute of Physics, Chinese Academy of Sciences, Beijing 100190, China \\\\
$^2$Max Planck Institute for Chemical Physics of Solids, 01187 Dresden, Germany
}%

\date{\today}

\begin{abstract}
A large, strongly temperature-dependent Nernst coefficient, $\nu$, is observed between $T$ = 2 K and 300 K for CeCu$_2$Si$_2$ and Ce$_{0.8}$La$_{0.2}$Cu$_2$Si$_2$. The enhanced $\nu(T)$ is determined by the asymmetry of the  on-site Kondo (conduction electron$-$4$f$ electron) scattering rate. Taking into account the measured Hall mobility, $\mu_H$, the highly unusual thermopower, $S$, of these systems can be semiquantitatively described  by $S(T)$ $=$ $-$$\nu(T)$/$\mu_H(T)$, which explicitly demonstrates that the thermopower originates from the local Kondo scattering process over a wide temperature range from far above to well below the coherence temperature ($\approx$ 20 K for CeCu$_2$Si$_2$). Our results suggest that the Nernst effect can act as a proper probe of local charge-carrier scattering. This promises an impact on exploring the unconventional enhancement of the thermopower in correlated materials suited for potential applications.

\end{abstract}

\pacs{Valid PACS appear here}
\maketitle

Strongly correlated electron systems in various material classes like $d$- and $f$-electron-based intermetallics, organic charge-transfer salts and transition-metal oxides have been in the focus of condensed-matter research for the past three decades. These systems give rise to a number of intriguing emerging phenomena, such as high-$T_c$ superconductivity \cite{Bednorz86, hosono}, colossal magnetoresistance \cite{CMR} and heavy-fermion (HF) behavior \cite{Ott75}, including HF superconductivity \cite{Steglich79}. Apart from the fundamental interest, it is the potential for technical applications which has been fueling the exploration of strongly correlated materials.

Because of their giant thermoelectric power (TEP), Kondo systems are considered especially promising for thermoelectric applications at cryogenic temperatures \cite{mahan}. In the case of dilute Kondo alloys, e.g., (\underline{La},Ce)Al$_2$ \cite{CeAl2}, the giant TEP derives from the on-site scattering of the conduction electrons by the local 4$f$-electrons \cite{maekawa,Zlatic05,Zlatic07}. For dense Kondo or HF systems on the other hand, the enhanced TEP is commonly ascribed to a weakly dispersive band of heavy quasiparticles, caused by the coherent hybridization between the 4$f$-electrons and the conduction bands \cite{miyake05,Behnia04}.  This view is strictly correct only at sufficiently low temperatures, $T$ $\ll$ $T_{coh}$, the temperature below which phase coherence develops in the Kondo scattering. On the other hand, research aiming at thermoelectric applications focuses on the giant TEP at $finite$ temperatures. In this case, a devoted theoretical treatment must account not only for the Kondo scattering, but also for the crystal-electric-field (CEF) splitting of the 4$f$ states and their hybridization with the conduction electrons \cite{Zlatic05,Zlatic07}.

In this letter we present an experimental study of the Nernst effect combined with Hall-effect measurements, which allow for a semiquantitative description of the complex TEP behavior of a canonical HF metal in a wide temperature range, from room temperature to $T$\,$<$\,$T_{coh}$. For this study, we have chosen the stoichiometric HF compound CeCu$_2$Si$_2$ and its 20at\% La-doped variant. Their TEP exhibits a broad positive peak at $T$ $\approx$ 160 K, followed by a pronounced negative one at $T$ $\approx$ 20 K and $\approx$ 12 K, respectively \cite{Franz78,Jaccard85,Ocko01}, see Fig.\,1b. These subsequent extrema are reflecting the single-ion Kondo temperatures (i) $T_{K,high}$ of the $J$ $=$ 5/2 Hund's rule multiplet state of Ce$^{3+}$, involving higher CEF states, and (ii) $T_K$ of the CEF doublet ground state \cite{Coqblin76}. $T_K$ obtained from the TEP agrees well with the value estimated from the molar electronic entropy ($S_e$ $\approx$ $R$ln2). It also coincides with the position of the low-temperature peak in the electrical resistivity curve, $\rho(T)$, which is a good empirical measure of the coherence temperature, i.e., $T_{coh}$ $\approx$ 20 K for CeCu$_2$Si$_2$, see inset of Fig. 2. Such a concurrence of $T_K$ and $T_{coh}$ has already been stated for CeCu$_2$Si$_2$ \cite{Oliver11} as well as the isostructural HF metals YbRh$_2$Si$_2$ \cite{Ernst11} and CeNi$_2$Ge$_2$ \cite{pikul12}. For comparison, we also present results on the nonmagnetic reference compound LaCu$_2$Si$_2$.

 Polycrystalline Ce$_{1-x}$La$_x$Cu$_2$Si$_2$ samples with $x$ = 0, 0.2 and 1.0 were prepared as described elsewhere \cite{Ocko01}. The measurements of the TEP $S(T)$, Nernst coefficient $\nu(T)$ and Hall coefficient $R_H(T)$ were carried out between 2 K and room temperature. For the Nernst effect and TEP measurements, we employed one chip resistor (2000\,$\Omega$) as heater and one thin ($\phi$ $=$ 25\,$\mu$m) chromel-AuFe$_{\rm 0.07\%}$ thermocouple for detecting the temperature gradient. The Nernst coefficient at each fixed temperature was measured in magnetic fields applied in opposite directions to cancel out the longitudinal TEP component. This procedure was further repeated at least five times to improve the resolution. No obvious magnetic-field dependence of $\nu(T)$ is observed up to 7 T and down to 2 K for the samples under investigation.

\begin{figure}[t]
\includegraphics[width=0.7\linewidth]{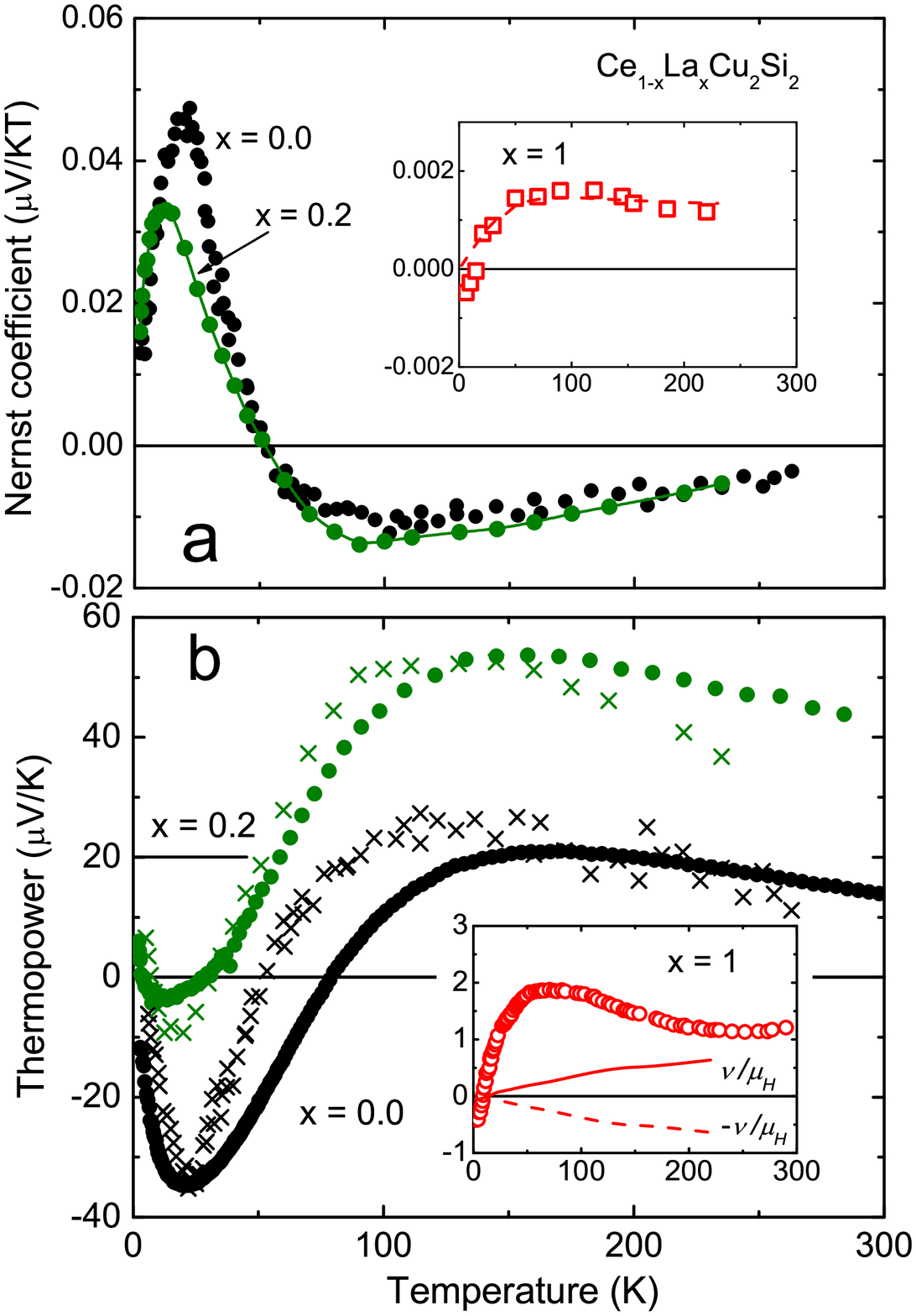}
\caption{(a): Nernst coefficient $\nu(T)$ of CeCu$_2$Si$_2$, Ce$_{0.8}$La$_{0.2}$Cu$_2$Si$_2$ (main panel) and LaCu$_2$Si$_2$ (inset). Dashed line in the inset was calculated for LaCu$_2$Si$_2$ under the assumption of dominating acoustic-phonon scattering, with $\tau(\epsilon)$\,$\propto$\,$\epsilon^r$ and $r$\,$=$\,$-\frac{1}{2}$, cf. Eq.\,2 \cite{Sond}. (b): Thermopower $S(T)$ compared to the ratio $-\nu/\mu_H$ vs $T$ (crosses) for the same compounds as in (a). For Ce$_{0.8}$La$_{0.2}$Cu$_2$Si$_2$ the vertical axis is shifted upward by 20 $\mu$V/K for the sake of clarity. In contrast to the cases of CeCu$_2$Si$_2$ and Ce$_{0.8}$La$_{0.2}$Cu$_2$Si$_2$, no agreement is observed for LaCu$_2$Si$_2$ between $\pm\,(\nu/\mu_H$) vs $T$ and $S(T)$ (inset). Axes in the insets are in the same units used for the respective main panels.
\label{nus.eps}}
\end{figure}

As displayed in Fig.\,1a, the Nernst coefficient $\nu(T)$  is similar for CeCu$_2$Si$_2$ and Ce$_{0.8}$La$_{0.2}$Cu$_2$Si$_2$. It exhibits a broad negative peak at $T$ $\approx$ 100 K, and a positive peak at 20 and 12 K, respectively. By contrast, $\nu(T)$ of LaCu$_2$Si$_2$ (inset of Fig.\,1a) shows a completely different temperature profile, with its values less than 2\,nV/KT over the whole $T$-range measured, spanning vertically a window that is a factor of 30 smaller than for CeCu$_2$Si$_2$. Note that $\nu(T)$ of the two Ce-based systems bears large resemblance to their TEP vs temperature results (cf. Fig.\,1b), except for the opposite sign \cite{sign}. The complex TEP behavior as observed in CeCu$_2$Si$_2$ has been explained by Zlati\'c $et$ $al.$ \cite{Zlatic05,Zlatic07} who consider the renormalized $f$-electron spectral weight distribution as a function of temperature in response to the interplay of multiple energy scales, namely the single-ion Kondo effect, the CEF splitting and the strength of the hybridization of the conduction band with the local 4$f$-states.

Following the description given by Mott \cite{Mott}, the thermoelectric tensor is determined by the logarithmic energy derivative of the electrical conductivity tensor. For a degenerate electron system, the diagonal term, i.\,e., the TEP is described by
\begin{equation}
S\,=\,-\frac{\pi^2}{3}\frac{k_{\rm B}^2 T}{e}[\frac{\partial {\rm ln  \tau }}{\partial \epsilon} + \frac{\partial {\rm ln}N} {\partial \epsilon}] _{\epsilon = \epsilon _F},
\end{equation}
with $\tau(\epsilon)$ denoting the relaxation time of the conduction electrons, and $N(\epsilon)$ denoting the electronic density of states (DOS) at the Fermi level $\epsilon _F$. The scattering events usually encountered involve a power-law dependence of the scattering time $\tau(\epsilon)$\,$\sim$\,$\epsilon^r$, with $r$ $\approx$ 1, which implies a weak energy dependence of $\tau(\epsilon)$. Therefore, one of the prevailing strategies for designing efficient thermoelectric materials is to search for a largely asymmetric DOS at $\epsilon _F$. This may be realized in materials with complex band structure and/or significant correlation effects \cite{mahan}. The first term in Eq.\,1, which is due to asymmetric scattering, has been practically ignored in most experimental and theoretical investigations, particularly on conventional, i.e., uncorrelated or only weakly correlated systems.

Like its diagonal counterpart (TEP), the off-diagonal terms of the thermoelectric tensor define the Nernst coefficient, which can be expressed by the derivative of the Hall angle tan$\theta _H$ \cite{Behnia,wang01,Sond}:
\begin{equation}
\nu\,=\,-\frac{\pi^2}{3}\frac{k_{\rm B}^2 T}{e}\frac{1}{B}\frac{\partial {\rm tan} \theta _H}{\partial \epsilon} | _{\epsilon = \epsilon _F}.
\end{equation}
The Nernst effect is sensitive to the details of the scattering processes in which the conduction electrons are involved. For example, for a simple solid with a single conduction band, tan$\theta _H$ is proportional to $\tau$, tan$\theta _H$ = $eB\tau/m^*$ = $\mu_HB$, where $B$ denotes the applied magnetic field, $m^*$ the effective mass, and $\mu_H$ (defined as $R_H$/$\rho$) the Hall mobility of the charge carriers \cite{Seeger,Behnia}. The majority of Nernst investigations in the literature have focused on semiconductors (showing a relatively large response), in order to figure out the dominating scattering mechanism for the charge carriers \cite{Seeger}. Except for the cuprates \cite{wang01},  the Nernst effect has been generally believed to be negligibly small in metallic systems \cite{Sond}.

According to Eqs.\,1-2, the Nernst coefficient $\nu$ would naturally be linked to the first term in Eq.\,1, as long as tan$\theta _H$ is related to $\tau$. In a Kondo system, it is known that the skew-scattering derived anomalous Hall coefficient $R_H$\,=\,$\xi$$\rho_{mag}$$\tilde{\chi}$ \cite{Coleman,Fert}, with $\xi$ being a parameter determined by the phase shift of the Kondo scattering, $\rho_{mag}$ the magnetic contribution to resistivity and $\tilde{\chi}$ the reduced susceptibility. Within the single-impurity Anderson model, $\tilde{\chi}$ is identical to the renormalized $f$-electron DOS $N_f$ \cite{Coleman} and $N_f$ $\propto$ $1/\tau$, i.e., the conduction-electron scattering rate \cite{Zlatic05}. It is, therefore, expected that tan$\theta_H$\,$\propto$\,$1/\tau$ in a Kondo system, assuming $\rho_{mag}$ dominates the measured $\rho$. Combining Eqs.\,1 and 2 and adopting the different relationships between tan$\theta _H$ and $\tau$ as mentioned above, one obtains
\begin{equation}
S\,=\,\pm\frac{\nu}{\mu _H} + (-\frac{\pi^2}{3} \frac{k_{\rm B}^2 T}{e} \frac{\partial {\rm ln}N} {\partial \epsilon}|_{\epsilon = \epsilon _F}),
\end{equation}
with the sign $+$ and $-$ before ${\nu}/{\mu _H}$ corresponding to tan$\theta_H$\,$\propto$\,$\tau$ (simple one-band metal) and tan$\theta_H$\,$\propto$\,$1/\tau$ (Kondo system), respectively. Note that, in deriving Eq.\,3, the detailed function relating tan$\theta_H$ and $\tau$ is not needed. Clearly, by measuring both $\nu(T)$ and $\mu_H(T)$, Eq.\,3 enables one to seperate the ``scattering" contribution from the ``band" contribution to the TEP (cf. Eq.\,1). Labelling the latter term $S_{\rm \small DOS}$ yields
\begin{equation}
\nu = \pm (S - S_{\rm DOS})\, {\rm tan} \,\theta _H/B,
\end{equation}
which explicitly explains the so-called Sondheimer cancellation in the Nernst effect of simple metals \cite{Sond,wang01}: the band term $S_{\rm DOS}$, which is usually dominating the TEP, is cancelled out in a Nernst measurement. Consequently, in simple metals, $\nu$ is almost negligible.

\begin{figure}[t]
\includegraphics[width=0.95\linewidth]{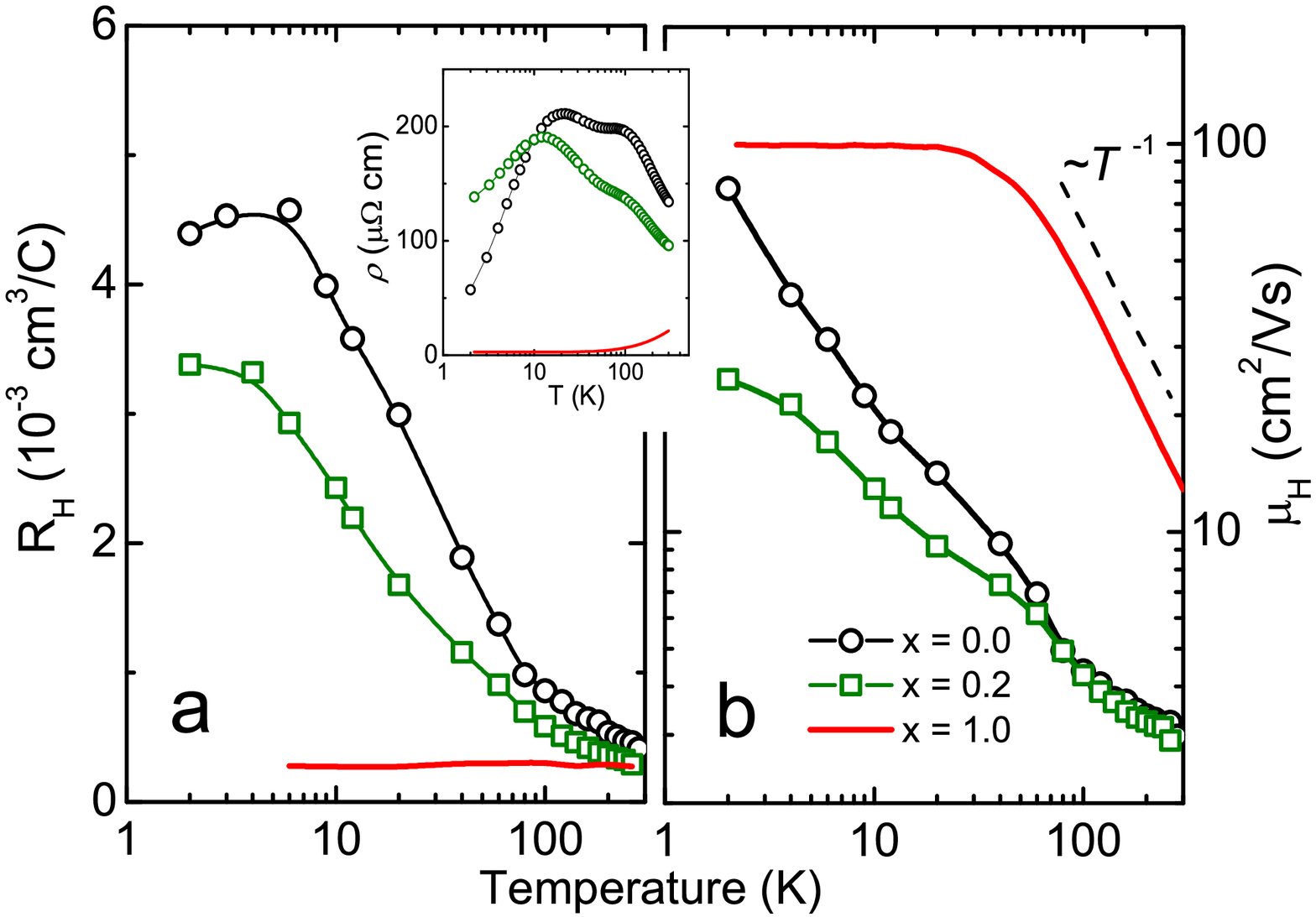}
\caption{Hall coefficient $R_H(T)$ in a semi-log plot (a) and Hall mobility $\mu _H(T)$ in a double-log plot (b) for CeCu$_2$Si$_2$,  Ce$_{0.8}$La$_{0.2}$Cu$_2$Si$_2$ and LaCu$_2$Si$_2$. Inset displays the temperature dependence of the electrical resistivity for the same compounds in a semi-log representation.
\label{nus.eps}}
\end{figure}

We now turn to the electrical transport properties of CeCu$_2$Si$_2$ and Ce$_{0.8}$La$_{0.2}$Cu$_2$Si$_2$, in comparison to the ones of the nonmagnetic reference system LaCu$_2$Si$_2$. For the two Ce-based systems, the Hall coefficient $R_H(T)$ (Fig.\,2a) is positive and increases with decreasing temperature, approaching a shallow maximum at $T$\,$<$\,5 K. By contrast, $R_H(T)$ of LaCu$_2$Si$_2$ shows a small positive, nearly constant value over the whole temperature range, indicating a dominating hole pocket in the valence band. The Hall mobility $\mu _H(T)$ for both CeCu$_2$Si$_2$ and Ce$_{0.8}$La$_{0.2}$Cu$_2$Si$_2$ (Fig.\,2b) increases monotonically upon decreasing temperature. Following the discussion by Coleman $et$ $al.$ \cite{Coleman} as well as Fert and Levy \cite{Fert}, this indicates a monotonic increase of $N_f$ (or $1/\tau$) and is characteristic of the local Kondo scattering.

As displayed in Fig.\,1b, $-\nu/ \mu _H$ vs $T$ (crosses) shows reasonable agreement, concerning both magnitude and sign, with the measured $S(T)$ over the whole temperature range for both CeCu$_2$Si$_2$ and Ce$_{0.8}$La$_{0.2}$Cu$_2$Si$_2$, an observation which is in line with tan$\theta_H$\,$\propto$\,$1/\tau$ expected for a Kondo system. Here, solely the scattering term in Eq.\,3, $-\nu(T)/\mu_H(T)$, is apt to describe reasonably well the measured TEP in wide range of temperature, i.e., from room temperature to $T$\,$<$\,$T_{coh}$.  As already mentioned, $T_{coh}$ is taken from the position of the low-$T$ peak in $\rho(T)$ (inset of Fig.\,2) which overlaps with a positive maximum in $\nu(T)$ (Fig.\,1a) and a negative one in $S(T)$ (Fig.\,1b).  The absence of a significant contribution of $S_{\rm DOS}$ to the TEP highlights the importance of the local Kondo scattering process over a large temperature range above 2\,K in CeCu$_2$Si$_2$ and Ce$_{0.8}$La$_{0.2}$Cu$_2$Si$_2$. This parallels the observation by Aliev $et$ $al.$ \cite{Aliev83} that the Hall effect of CeCu$_2$Si$_2$ is strongly enhanced upon cooling from far above to well below $T_{coh}$, which they ascribed to the formation of the local Kondo resonance. This conclusion is strongly supported by Coleman $et$ $al.$ \cite{Coleman}, who calculated the anomalous Hall effect in a Kondo-lattice system in terms of the onsite skew scattering within the framework of a local Fermi-liquid model.

The preceding analysis does not apply to LaCu$_2$Si$_2$: as seen in the inset of Fig.\,1b, for this compound, $\pm\nu(T)/\mu _H(T)$ cannot explain the measured $S(T)$. Assuming that the dominating electron scattering in LaCu$_2$Si$_2$ is due to acoustic phonons, with the scattering exponent $r$ $=$ $-$$\frac{1}{2}$, one can calculate the expected $\nu(T)$ for this material, following Eq.\,2 \cite{Sond}. Indeed, this calculation (dashed line, inset of Fig.\,1a) yields a good agreement with the measured $\nu(T)$ above 30 K. It is consistent with the carrier mobility $\mu_H(T)$ (Fig.\,2b), which exhibits a $T^{-1}$ dependence above 50 K, characteristic of conduction-electron scattering in a metal being dominated by acoustic phonons.
Therefore, the band term $S_{\rm DOS}$ contributes significantly to the TEP of LaCu$_2$Si$_2$, presumably along with a phonon-drag term which partly adds to the broad $S(T)$ hump at $T$ $\approx$ 50 K.

Successfully mapping the measured $S(T)$ by $\nu(T)$ (divided by $\mu_H(T)$) in the two Ce-based systems over a wide $T$ range reveals that both quantities can be described solely by a single relaxation-time $\tau(\epsilon)$ of the conduction electrons. Apparently, this is based on the fact that the anomalous Hall effect in a Kondo lattice is due to the formation of the Kondo resonance, which guarantees that the Hall angle can be described by $N_f$\,($\propto$\,$1/\tau$) down to well below $T_{coh}$ \cite{Coleman}.
The latter case is totally different from that of, e.g., a magnetic semiconductor, where a huge Nernst coefficient has also been observed \cite{Pu}. In such a system, no connection can be built among $S$ and $\nu$, because here the Nernst coefficient is dominated by skew scattering from local magnetic moments, where tan$\theta_H$ and $\tau$ are independent of each other. Recent measurements of $\nu(T)$, $S(T)$, $R_H(T)$, and $\rho(T)$ over an extended temperature range were reported by Matusiak $et$ $al.$ for the HF metal Ce$_2$PdIn$_8$ \cite{218}. There, the enhanced Hall coefficient $R_H(T)$, which is largely different from the typical behavior expected for Kondo scattering, is ascribed to antiferromagnetic spin fluctuations causing a highly anisotropic scattering time \cite{218}. For Ce$_2$PdIn$_8$, the Nernst coefficient has a similar temperature profile as $R_H(T)$, strikingly different from the canonical behavior of CeCu$_2$Si$_2$ discussed above.

To substantiate our conclusion that the Nernst effect is probing the single-ion Kondo effect, we refer to Zlati\'c $et$ $al.$ \cite{Zlatic07}, who calculated the thermoelectric response of a periodic Anderson model by considering the on-site Kondo scattering. This way, they showed that the ``band" term, ($\partial/\partial\epsilon$)(ln$N_c$)$|_{\epsilon = \epsilon _F}$,
due to the {\it renormalized conduction-band DOS}, $N_c$, is negligibly small compared to the scattering term. The latter is related to the {\it renormalized $f$-electron DOS}, $N_f$, by
\begin{equation}
\frac{\partial}{\partial\epsilon}[{\rm ln}\tau(\epsilon)] | _{\epsilon = \epsilon _F} \simeq \pm \frac{2LN_f(\epsilon_F)Z_f^{-1}}{n_c},
\end{equation}
where $L$ denotes the degeneracy of $f$-electron spin state, $Z_f$ the renormalization factor, $N_f$ the renormalized $f$-electron DOS, $n_c$ the number of conduction electrons per site. Eq.\,5 explicitly links the asymmetric Kondo scattering of conduction electrons to the renormalized $f$ spectral weight, and microscopically explains that, while the TEP of the Kondo lattice is determined by the scattering approach in a wide range of temperature, it is adequately described by the huge quasiparticle DOS at even lower temperatures, $T$$\ll$$T_{coh}$.  In addition, the ``energy resolution" of the Nernst measurement, determined by the applied temperature difference $\delta$$T$ during the measurement, has also to be considered. Because we applied a $\delta T$ $<$ 0.02\,$T$ (for example, $\delta$$T$ $<$ 0.2 K at $T$ $=$ 10 K) in our measurements, $\delta T$ is much smaller than the Kondo temperature ($T_K$ $\approx$ 20 K for CeCu$_2$Si$_2$), which determines the energy width of the Kondo resonance. This makes it possible for the Nernst effect to resolve the asymmetry of the Kondo scattering.

In conclusion, we showed that the enhanced TEP as well as the Nernst coefficient of the Kondo-lattice system CeCu$_2$Si$_2$ and its 20 at\% La-doped variant can be well explained by the highly asymmetric Kondo scattering process of conduction electrons, without resorting to the electronic structure of the heavy quasiparticles. This strongly indicates that the single-ion Kondo physics applies well to the Kondo-lattice system CeCu$_2$Si$_2$ down to $T$\,$<$\,$T_{coh}$. A similar conclusion was recently drawn for CeNi$_2$Ge$_2$ from calorimetric results on the quasibinary alloys (Ce$_{1-x}$La$_x$)Ni$_2$Ge$_2$ \cite{pikul12}. Our results indicate the Nernst effect to be a promising probe for directly detecting asymmetric scattering processes in correlated-electron systems and, therefore, to play an important role in exploring unconventional thermoelectric materials for potential applications.

The authors thank S. Wirth and S. Kirchner for most valuable comments on this manuscript and C. Geibel for providing the samples. P.S. thanks financial support from the MOST of China (Grant No: 2012CB921701). F.S. acknowledges partial support by the DFG through FG960 ``Quantum Phase Transitions".


\begin{thebibliography}{99}
\bibitem{sign} We adopt the sign convention commonly used in published work on cuprates, such that the Nernst effect arising from vortex flow is positive.
\end{thebibliography}
\end{document}